\pdfoutput=1
\documentclass[aps,prl,preprint,superscriptaddress,tightenlines,nofootinbib,showpacs]{revtex4-1}

\usepackage{graphicx}
\usepackage{dcolumn}
\usepackage{bm}
\usepackage{hyperref}
\usepackage{amssymb,amsmath}
\IfFileExists{txfonts.sty}{\usepackage{txfonts}}{}
\usepackage{xspace}
\xspaceaddexceptions{]}

\newcommand{\Dspm}{\ensuremath{D_s^\pm}}

\newcommand{\Ds}{\ensuremath{D_{s}}\xspace}
\newcommand{\Dsstar}{\ensuremath{\Ds^*}\xspace}
\newcommand{\Dsp}{\ensuremath{D_{s}^+}\xspace}
\newcommand{\Dsm}{\ensuremath{D_{s}^-}\xspace}

\newcommand{\mrec}{\ensuremath{m_\mathrm{rec}}\xspace}

\newcommand{\GeVcsq}{GeV\ensuremath{/c^2}\xspace}
\newcommand{\MeVcsq}{MeV\ensuremath{/c^2}\xspace}
\newcommand{\MeVc}{MeV\ensuremath{/c}\xspace}
\newcommand{\GeVc}{GeV\ensuremath{/c}\xspace}

\newcommand{\Dz}{\ensuremath{D^0}}

\newcommand{\Dp}{\ensuremath{D^+}}

\newcommand{\Kp}{\ensuremath{K^+}}
\newcommand{\Km}{\ensuremath{K^-}}
\newcommand{\KS}{\ensuremath{K_{S}^0}\xspace}
\newcommand{\pip}{\ensuremath{\pi^+}}
\newcommand{\pim}{\ensuremath{\pi^-}}
\newcommand{\piz}{\ensuremath{{\pi^0}}}

\newcommand{\invpb}{pb$^{-1}$}

\newcommand{\Br}{\ensuremath{\mathcal{B}}}
\newcommand{\cleoc}{\hbox{CLEO-c}\xspace}
\newcommand{\etatpi}{\ensuremath{\eta_{3\pi}}}
\newcommand{\etagg}{\ensuremath{\eta_{\gamma\gamma}}}
\newcommand{\etaprgg}{\ensuremath{\eta'_{\gamma\gamma}}}
\newcommand{\etaprtpi}{\ensuremath{\eta'_{3\pi}}}
\newcommand{\etaprrg}{\ensuremath{\eta'_{\rho\gamma}}}
\newcommand{\com}{\ensuremath{\mathrm{cm}}}

\begin{document}
\pdfpageheight\paperheight
\pdfpagewidth\paperwidth


\title{\boldmath Improved Measurement of Absolute Hadronic Branching Fractions of the \Dsp\ Meson}
\preprint{CLNS 13/2086}  
\preprint{CLEO 13-01}    

\author{P.~U.~E.~Onyisi}
\affiliation{University of Texas at Austin, Austin, Texas 78712, USA}
\author{G.~Bonvicini}
\author{D.~Cinabro}
\author{M.~J.~Smith}
\author{P.~Zhou}
\affiliation{Wayne State University, Detroit, Michigan 48202, USA}
\author{P.~Naik}
\author{J.~Rademacker}
\affiliation{University of Bristol, Bristol BS8 1TL, United Kingdom}
\author{K.~W.~Edwards}
\affiliation{Carleton University, Ottawa, Ontario, Canada K1S 5B6}
\author{R.~A.~Briere}
\author{H.~Vogel}
\affiliation{Carnegie Mellon University, Pittsburgh, Pennsylvania 15213, USA}
\author{J.~L.~Rosner}
\affiliation{University of Chicago, Chicago, Illinois 60637, USA}
\author{J.~P.~Alexander}
\author{D.~G.~Cassel}
\author{S.~Das}
\author{R.~Ehrlich}
\author{L.~Gibbons}
\author{S.~W.~Gray}
\author{D.~L.~Hartill}
\author{B.~K.~Heltsley}
\author{D.~L.~Kreinick}
\author{V.~E.~Kuznetsov}
\author{J.~R.~Patterson}
\author{D.~Peterson}
\author{D.~Riley}
\author{A.~Ryd}
\author{A.~J.~Sadoff}
\author{X.~Shi}
\altaffiliation[Now at: ]{National Taiwan University, Taipei, Taiwan}
\author{W.~M.~Sun}
\affiliation{Cornell University, Ithaca, New York 14853, USA}
\author{J.~Yelton}
\affiliation{University of Florida, Gainesville, Florida 32611, USA}
\author{P.~Rubin}
\affiliation{George Mason University, Fairfax, Virginia 22030, USA}
\author{N.~Lowrey}
\author{S.~Mehrabyan}
\author{M.~Selen}
\author{J.~Wiss}
\affiliation{University of Illinois, Urbana-Champaign, Illinois 61801, USA}
\author{J.~Libby}
\affiliation{Indian Institute of Technology Madras, Chennai, Tamil Nadu 600036, India}
\author{M.~Kornicer}
\author{R.~E.~Mitchell}
\affiliation{Indiana University, Bloomington, Indiana 47405, USA }
\author{D.~Besson}
\affiliation{University of Kansas, Lawrence, Kansas 66045, USA}
\author{T.~K.~Pedlar}
\affiliation{Luther College, Decorah, Iowa 52101, USA}
\author{D.~Cronin-Hennessy}
\author{J.~Hietala}
\affiliation{University of Minnesota, Minneapolis, Minnesota 55455, USA}
\author{S.~Dobbs}
\author{Z.~Metreveli}
\author{K.~K.~Seth}
\author{A.~Tomaradze}
\author{T.~Xiao}
\affiliation{Northwestern University, Evanston, Illinois 60208, USA}
\author{A.~Powell}
\author{C.~Thomas}
\author{G.~Wilkinson}
\affiliation{University of Oxford, Oxford OX1 3RH, United Kingdom}
\author{D.~M.~Asner}
\author{G.~Tatishvili}
\affiliation{Pacific Northwest National Laboratory, Richland, Washington 99352, USA}
\author{J.~Y.~Ge}
\author{D.~H.~Miller}
\author{I.~P.~J.~Shipsey}
\author{B.~Xin}
\affiliation{Purdue University, West Lafayette, Indiana 47907, USA}
\author{G.~S.~Adams}
\author{J.~Napolitano}
\affiliation{Rensselaer Polytechnic Institute, Troy, New York 12180, USA}
\author{K.~M.~Ecklund}
\affiliation{Rice University, Houston, Texas 77005, USA}
\author{J.~Insler}
\author{H.~Muramatsu}
\author{L.~J.~Pearson}
\author{E.~H.~Thorndike}
\affiliation{University of Rochester, Rochester, New York 14627, USA}
\author{M.~Artuso}
\author{S.~Blusk}
\author{R.~Mountain}
\author{T.~Skwarnicki}
\author{S.~Stone}
\author{J.~C.~Wang}
\author{L.~M.~Zhang}
\affiliation{Syracuse University, Syracuse, New York 13244, USA}
\collaboration{CLEO Collaboration}
\noaffiliation

\def\usesections{0}

\date{\today}

\begin{abstract} 
The branching fractions of \Dspm\ meson decays serve to normalize many measurements of processes involving charm quarks.  Using 586~\invpb\ of $e^+ e^-$ collisions recorded at a center of mass energy of 4.17~GeV, we determine absolute branching fractions for 13 \Dspm\ decays in 16 reconstructed final states with a double tag technique.  In particular we make a precise measurement of the branching fraction $\Br(\Ds \to \Km\Kp\pip) = (5.55 \pm 0.14 \pm 0.13)\%$, where the uncertainties are statistical and systematic respectively.  We find a significantly reduced value of $\Br(\Ds\to\pip\piz\eta')$ compared to the world average, and our results bring the inclusively and exclusively measured values of $\Br(\Ds\to\eta' X)$ into agreement.  We also search for $CP$-violating asymmetries in $\Ds$ decays and measure the cross-section of $e^+ e^- \to \Ds^* \Ds$ at $E_\com = 4.17$~GeV.
\end{abstract}

\maketitle

\if\usesections1\section{Introduction}\fi

Measurements of absolute hadronic branching fractions of ground state charmed mesons are important for several reasons.  The branching fractions for certain decays, such as $\Dsp \to \Km\Kp\pip$ or $\Dz\to\Km\pip$, serve to normalize measurements of decay chains involving charm quarks.  Understanding $\Dspm$ decays is particularly important for studies of the $B_s^0$ meson, the decays of which are dominated by final states involving $\Dspm$ \cite{Beringer:1900zz}.  In addition, hadronic decays probe the interplay of short distance weak decay matrix elements and long distance QCD interactions, and measurements of branching fractions provide valuable information to help understand strong force-induced amplitudes and phases \cite{Bhattacharya:2008ss,*Bhattacharya:2008ke,*Cheng:2010ry,*Fusheng:2011tw}.


The \cleoc experiment at the Cornell Electron Storage Ring (CESR) $e^+ e^-$ collider collected $586 \pm 6$~\invpb\ of data at a center of mass energy of 4.17 GeV, above the threshold for \Dsp\Dsm\ production, but below threshold for $\Dspm D K$.  As a consequence any event that contains a \Dsp\ meson also contains a \Dsm.  This condition enables the use of a powerful ``double tag'' technique for obtaining absolute branching fractions, pioneered by the Mark-III Collaboration \cite{Baltrusaitis:1985iw} and used in previous \cleoc determinations of branching fractions of \Dz, \Dp, and \Dsp\ decays \cite{Dobbs:2007ab,Alexander:2008aa}.  

The CLEO Collaboration has previously reported measurements of eight absolute \Dsp\ branching fractions with a 298~\invpb\ subset of the data \cite{Alexander:2008aa}.  In this paper we report the results of an extended determination of the branching fractions with the full \cleoc dataset, using 13 \Ds\ decays reconstructed in 16 final states, listed in Table~\ref{tbl:modes}.  This update significantly improves both the statistical and systematic uncertainties on the branching fraction determinations of key normalization modes.  We also obtain first measurements of the branching fractions of a number of previously unmeasured decays, and resolve the tension between the world average inclusive and exclusive determinations of $\Br(\Ds \to \eta' X)$.

\if\usesections1\section{Technique}\fi

Consider a situation in which $e^+ e^-$ collisions have produced a number $N_{\Ds\Ds}$ of \Dsp\Dsm\ pairs.  For each \Ds\ decay mode considered, there is a branching fraction $\Br_i \equiv \Br(\Dsp \to i)$.  We assume no $CP$ violation while determining the branching fractions, so $\Br(\Dsp \to i) = \Br(\Dsm \to \bar\imath)$.  We have 
\[ Y_i^+ = N_{\Ds\Ds} \Br_i \epsilon_i^+;\hspace{1em} Y_{\bar\jmath}^- = N_{\Ds\Ds} \Br_j \epsilon_{\bar\jmath}^-; \hspace{1em} Y_{i\bar\jmath} = N_{\Ds\Ds} \Br_i \Br_j \epsilon_{i\bar\jmath} \]
where $\epsilon_i^+$ is the efficiency for the detection of the decay $\Dsp \to i$ (a \textit{single tag}), $\epsilon_{\bar\jmath}^-$ is the efficiency for the detection of the decay $\Dsm \to \bar\jmath$ (a different single tag), $\epsilon_{i\bar\jmath}$ is the efficiency for the simultaneous detection of the two decays $\Dsp \to i$, $\Dsm \to \bar\jmath$ (a \textit{double tag}), and $Y_i^+$, $Y_{\bar\jmath}^-$, and $Y_{i\bar\jmath}$ are the yields for the two single tags and the double tag.  The efficiencies are determined from Monte Carlo simulations.  We determine $Y_i^+$ and $Y_i^-$, and $\epsilon_i^+$ and $\epsilon_i^-$, separately for each mode $i$.  For $M$ different final states, there are $2M$ single tag yields and $M^2$ double tag yields, leading to $2M+M^2$ relations and $M+1$ parameters ($\Br_i$ and $N_{\Ds\Ds}$) to determine.  For $M\ge 1$, this is an overconstrained system and allows us to determine the parameters via a likelihood fit.

We can test for direct $CP$ violation in \Dspm\ decays by computing the $CP$ asymmetries
\[ \mathcal{A}_{CP,i} = \frac{Y_i^+/\epsilon_i^+ - Y_{\bar\imath}^-/\epsilon_{\bar\imath}^-}{Y_i^+/\epsilon_i^+ + Y_{\bar\imath}^-/\epsilon_{\bar\imath}^-} \]
for each mode $i$.  The $\mathcal{A}_{CP}$ values do not depend on the branching fraction fit.

\begin{table}
\caption{\label{tbl:modes}The 16 \Ds final states used in this analysis. The symbols \etagg, \etatpi, \etaprgg, \etaprtpi, and \etaprrg\ are defined in the text.}
\begin{tabular}{lll}
\hline\hline
$\Dsp \to K_S^0 K^+$ &
$\Dsp \to K^- K^+ \pi^+$&
$\Dsp \to \KS \Kp \piz$ \\
$\Dsp \to \KS \KS \pip$ &
$\Dsp \to K^- K^+ \pi^+ \pi^0$ &
$\Dsp \to \KS \Kp \pip \pim$ \\
$\Dsp \to \KS \Km \pip \pip$ &
$\Dsp \to \pi^+ \pi^+ \pi^-$&
$\Dsp \to \pip \etagg$ \\
$\Dsp \to \pip \etatpi$ &
$\Dsp \to \pip \piz \etagg$ &
$\Dsp \to \pip \etaprgg$\\
$\Dsp \to \pip \etaprtpi$ & 
$\Dsp \to \pip \etaprrg$ &
$\Dsp \to \pip \piz \etaprgg$ \\
$\Dsp \to \Kp \pip \pim$\\
\hline\hline
\end{tabular}
\end{table}

\if\usesections1\section{\label{sec:reco}Detector and Particle Reconstruction}\fi
The \cleoc detector was a symmetric general purpose solenoidal particle detector located at the CESR $e^+ e^-$ collider.  The detector is described in detail elsewhere \cite{Viehhauser:2001ue,Briere:2001rn}.  Here we summarize the details relevant for this measurement.

The momenta of long-lived charged particles, in particular $\pi^\pm$ and $K^\pm$, are determined using two concentric drift chambers \cite{Peterson:2002sk,Briere:2001rn} immersed in a 1 T magnetic field.  The tracking system provides angular coverage in the region $|\cos\theta\,| < 0.93$, where $\theta$ is the polar angle from the beam axis, and has momentum resolution $\sigma_p/p \sim 0.6\%$ at 1 \GeVc\ for tracks that cross every layer.  Discrimination between different species of charged particles is achieved by using specific ionization ($dE/dx$) measurements from the outer drift chamber and particle velocity as measured by a ring-imaging Cherenkov detector \cite{Artuso:2000mw} for $|\cos \theta\,| < 0.8$. Photons are detected as showers in a CsI(Tl) electromagnetic calorimeter \cite{Kubota:1991ww}, which provides energy resolution of $\sigma_E/E \sim 5\%$ at 100 MeV.

Charged pions and kaons are selected from reconstructed charged tracks that satisfy $dE/dx$ and Cherenkov requirements.  The minimum track momenta considered in this analysis are 50 \MeVc\ and 125 \MeVc\ for $\pi^\pm$ and $K^\pm$ respectively.

We form \KS candidates from pairs of opposite sign charged tracks.  They are constrained to originate at a common origin which may be displaced from the primary collision vertex, and the four-momentum of the system is recomputed at that point.  We require $|m(\pip\pim) - 497.7\textrm{ MeV}/c^2| < 6.3$ \MeVcsq, where $m(\pip\pim)$ is the reconstructed invariant mass of the \pip\pim\ pair.

Pairs of photon candidates are combined to form $\piz \to \gamma\gamma$ and $\eta \to \gamma\gamma$ ($\eta_{\gamma\gamma}$) candidates and kinematic fits to the $\piz$ and $\eta$ masses are performed to improve the four-momentum resolution.  We require that the unconstrained masses be within $3\sigma$ of the nominal particle mass given the expected resolution.  We reconstruct $\eta \to \pip\pim\piz$ (\etatpi) candidates, requiring that $0.53 < m(\pip\pim\piz) < 0.57$ \GeVcsq.

We form $\eta'$ candidates in three final states: $\pip\pim\eta_{\gamma\gamma}$ (\etaprgg), $\pip\pim\etatpi$ (\etaprtpi), and $\pip\pim\gamma$ (\etaprrg).  The invariant mass of the reconstructed $\eta'$ candidate is required to satisfy $|m(\etaprgg)-957.8\textrm{ \MeVcsq}| < 10$ \MeVcsq, $|m(\etaprtpi) - 957.8\textrm{ \MeVcsq}| \le 15$ \MeVcsq, or $|m(\etaprrg) - 957.78\textrm{ \MeVcsq}| \le 18$ \MeVcsq.  For \etaprrg\ candidates, we additionally require $E_\gamma > 100$ MeV in the laboratory frame and $m(\pip\pim) > $ 0.5 \GeVcsq to reduce combinatoric background.

\if\usesections1\section{Event Selection}\fi
At the center of mass energy $E_\com = 4.17$ GeV, the cross-section for \Ds\ production is dominated by the process $e^+ e^- \to D_{s}^{*\pm} D_s^\mp \to \Dsp\Dsm(\gamma, \piz)$.  We refer to the \Ds\ meson that is a daughter of the $\Ds^*$ as the \textit{indirect} \Ds, and the other as the \textit{direct} \Ds.  We do not search for the $\gamma$ or \piz\ from the $\Ds^*$ decay.

\begin{figure}
 \includegraphics[width=.7\linewidth]{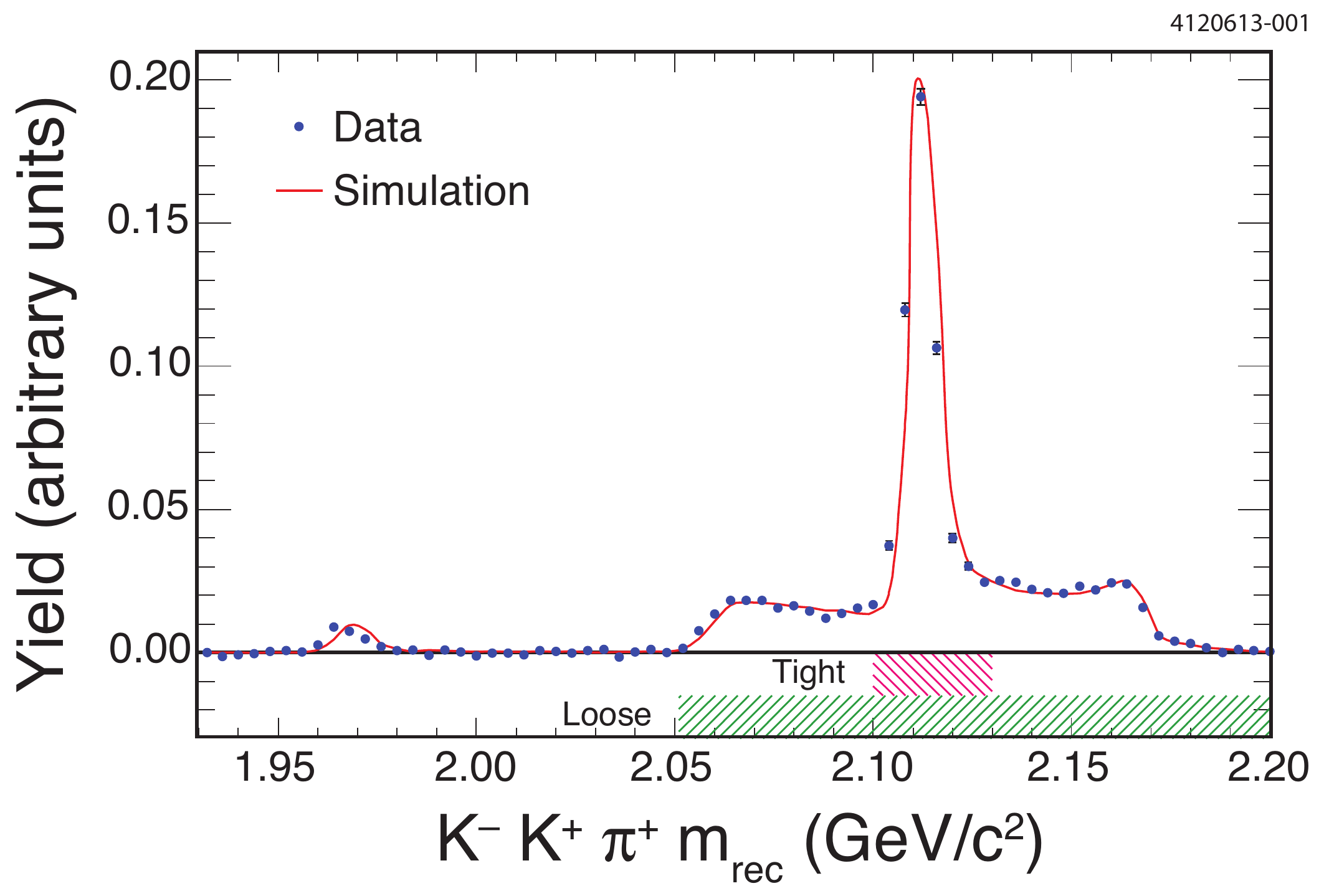}
 \caption{\label{fig:mrecoil}(Color online) Fitted yields of $\Dsp \to \Km\Kp\pip$ candidates in bins of recoil mass \mrec\ in data and simulation, and the tight and loose ranges used to select $e^+ e^- \to \Ds^{*\pm}\Ds^\mp$ events for further study.  The small peak at $\mrec \approx 1.97$ \GeVcsq is due to $e^+ e^- \to \Dsp\Dsm$.}
\end{figure}

The signal sample consists of single and double tagged events consistent with $e^+ e^- \to D_{s}^{*\pm} D_s^\mp$. This two-body production mode determines the momenta of direct \Ds mesons; indirect \Ds mesons have an additional momentum component from the $\Ds^*$ decay.  We define the recoil mass variable \mrec\ via
\[ \mrec^2 c^4 = \left(E_\com - \sqrt{\left|\mathbf{p}(\Ds)\right|^2 c^2 + m_{\Ds}^2 c^4}\right)^2 - \left|\mathbf{p}_\com - \mathbf{p}_{\Ds}\right|^2 c^2 \]
where $(E_\com, \mathbf{p}_\com)$ is the four-momentum of the colliding $e^+ e^-$ system and $\mathbf{p}(\Ds)$ is the measured momentum of the \Ds\ candidate.  The value of $m_{\Ds}$ is fixed at 1.9685 \GeVcsq.  The \mrec variable has a narrow peak at the \Dsstar mass for direct \Ds candidates and a broader distribution around the \Dsstar mass for indirect candidates.  The upper kinematic limit of \mrec, when the \Ds\ candidate has zero momentum, is $\approx 2.20$ \GeVcsq for $E_\com = 4.17$ GeV.  Both direct and indirect candidates are well separated in \mrec\ from \Ds\ candidates produced via $e^+ e^- \to \Dsp\Dsm$.  The distribution of \mrec\ for $\Dsp \to \Km\Kp\pip$ candidates is shown in Fig.~\ref{fig:mrecoil}.

For the $\KS\Kp$, $\Km\Kp\pip$, and $\pip\etaprgg$ single tag modes, we require a loose recoil mass cut $\mrec > 2.051$ \GeVcsq, which accepts both direct and indirect \Ds candidates.  For all other single tag modes, for greater background suppression, we require a tight cut $2.10 < \mrec < 2.13$ \GeVcsq which primarily accepts direct candidates.  These selections reject candidates from $e^+ e^- \to \Dsp\Dsm$.  In every event we search for all considered \Dsp and \Dsm single tag final states independently.  

All possible pairings of considered final states are searched for double tag candidates, giving $16^2 = 256$ modes.  In all double tag candidates one \Ds candidate should be direct and the other indirect; therefore we require one to have $\mrec > 2.1$ \GeVcsq and the other $\mrec > 2.051$ \GeVcsq.  Because this is looser than the tight single tag selection, it is possible for the \Ds\ candidates of a double tag not to be accepted as single tag candidates.  

We require that charged or neutral pions, including daughters of \KS, $\eta$, and $\eta'$ mesons, have momenta exceeding 100 \MeVc.  This removes the combinatorics associated with the large number of soft pions from $D^*$ decays.  In modes with exceptionally low background ($\KS \Kp$, $\Km\Kp\pip$, $\pip\etagg$, and $\pip\etaprgg$) this additional selection is not applied and the minimum pion momentum remains 50 \MeVc.

To remove contamination from \KS\ decays in the \pip\pip\pim\ and \Kp\pim\pip\ modes, candidates are vetoed if a pion pair satisfies 475 $< m(\pip\pim) <$ 520 \MeVcsq.

For $\pip\piz\etagg$, we enhance the signal by selecting only candidates where 670 $< m(\pip\piz) <$ 870 \MeVcsq\ (i.e., consistent with a $\rho^+$ decay).

Finally, a number of modes are contaminated by combinations of a real \Dz\ or \Dp\ and an additional pion.  For these modes we reject candidates where removing one pion leaves a system with invariant mass near that of the \Dz\ or \Dp.  This has negligible impact on signal candidates but simplifies the background description.

If there are multiple single tag candidates in an event for a given final state and charge, we choose the one with the smallest value of $|\mrec - 2.112\textrm{ GeV}/c^2|$.  Similarly, if there are multiple double tag candidates for a given mode in an event, the one with the average \Ds candidate invariant mass closest to 1.9682 \GeVcsq is chosen.  This resolution takes place after all other selections are applied.

\begin{figure*}
\begin{center}
 \includegraphics[width=\linewidth]{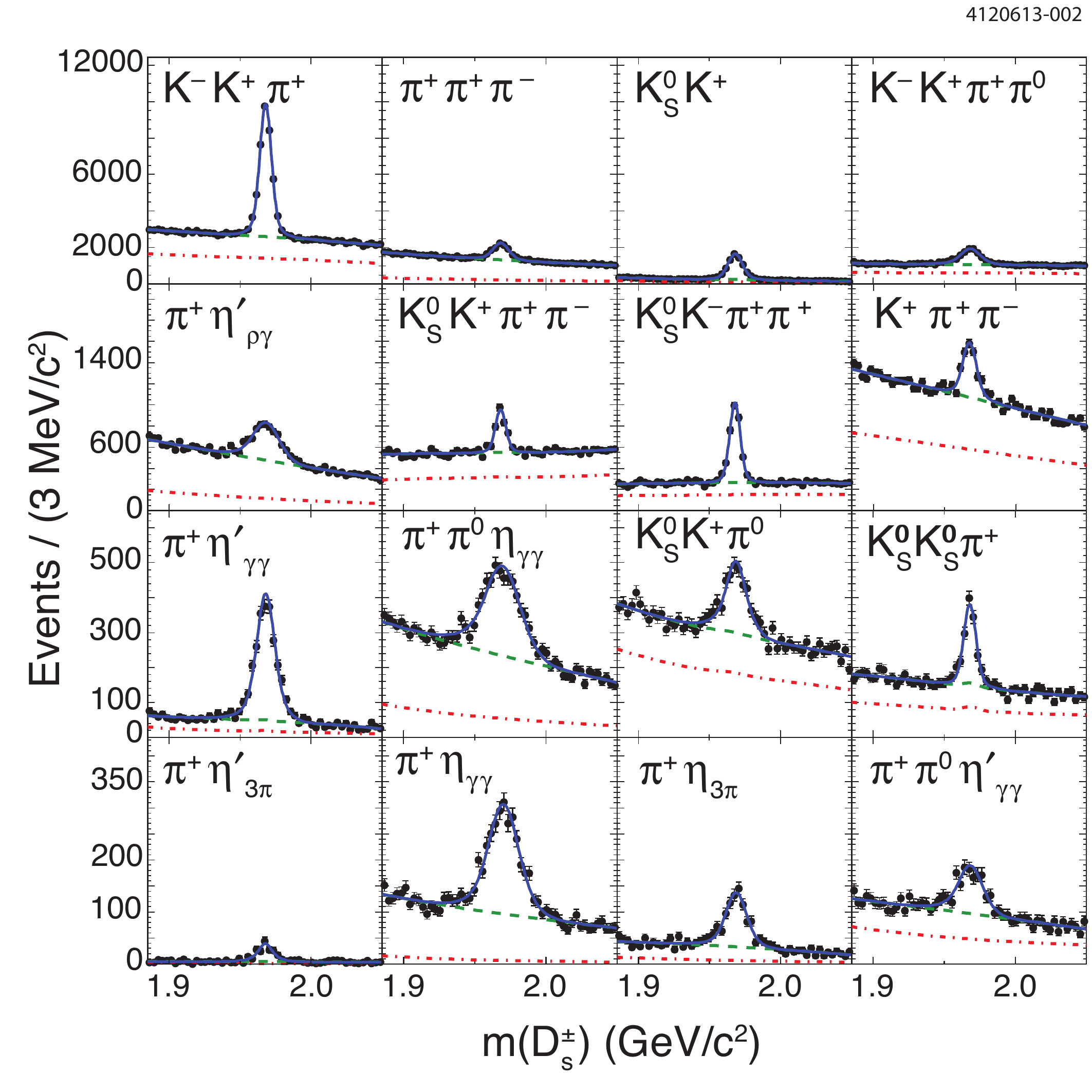}
 \caption{\label{fig:st}(Color online) Invariant mass spectra for single tags and corresponding yield fits for the sixteen reconstructed \Ds\ decay final states. Charge conjugate yields are combined in this figure.  The points are observed data; the blue solid lines are the best fit to a sum of background and signal.  The total background estimates are shown as the green dashed lines, while the background arising only from other open charm decays (which includes the peaking contributions) is shown as red dot-dashed lines.}
 \end{center}
\end{figure*}

\if\usesections1\section{Yields, Efficiencies, and Backgrounds}\fi
We fit the invariant mass spectrum of single tag candidates to obtain signal event yields; these fits are done separately for each mode and \Ds charge.  The (charge-combined) fits in data are shown in Fig.~\ref{fig:st}.  The background is parametrized as the sum of components from (a) other decays of charmed mesons (``open charm'') and (b) continuum light quark production, $\tau^+\tau^-$ production, and $\gamma J/\psi$ and $\gamma \psi (2S)$ production.  The open charm background often has significant structure, and the shapes are derived from a Monte Carlo simulation of inclusive open charm production processes (``generic MC").  The non-open charm background is parametrized by a quadratic polynomial whose parameters are allowed to float, which has been verified to be an acceptable model in Monte Carlo simulations of these processes.  

Open charm production and decay is modeled with the \textsc{EvtGen} package \cite{Lange:2001uf}, with decay tables tuned to reflect the \cleoc results for open charm branching fractions and production cross-sections at 4.17 GeV.  Initial 
state radiation is modeled using the cross-sections for open charm processes from threshold to the center 
of mass energy \cite{CroninHennessy:2008yi}.  Final state radiation from charged particles is modeled with the \textsc{Photos} 2.15 package \cite{Barberio:1990ms,Barberio:1993qi}.  Particle interactions with material and detector response are modeled with a \textsc{Geant} 3-based simulation \cite{geant}.  The normalization of the open charm background contribution is fixed from the generic MC prediction and the peaking component is generally negligible, although for modes with \KS\ mesons in the final state it can reach 5\% of the signal yield.

The signals for modes with photons are modeled with the sum of a Gaussian and a 
wider Crystal Ball function \cite{cbfunc} with a common mean parameter; all other modes are modeled with the sum of two Gaussians with a common mean.  The lineshapes and reconstruction efficiencies are determined from dedicated signal Monte Carlo samples (``signal MC''). 

\begin{figure}
\begin{center}
 \includegraphics[width=.7\linewidth]{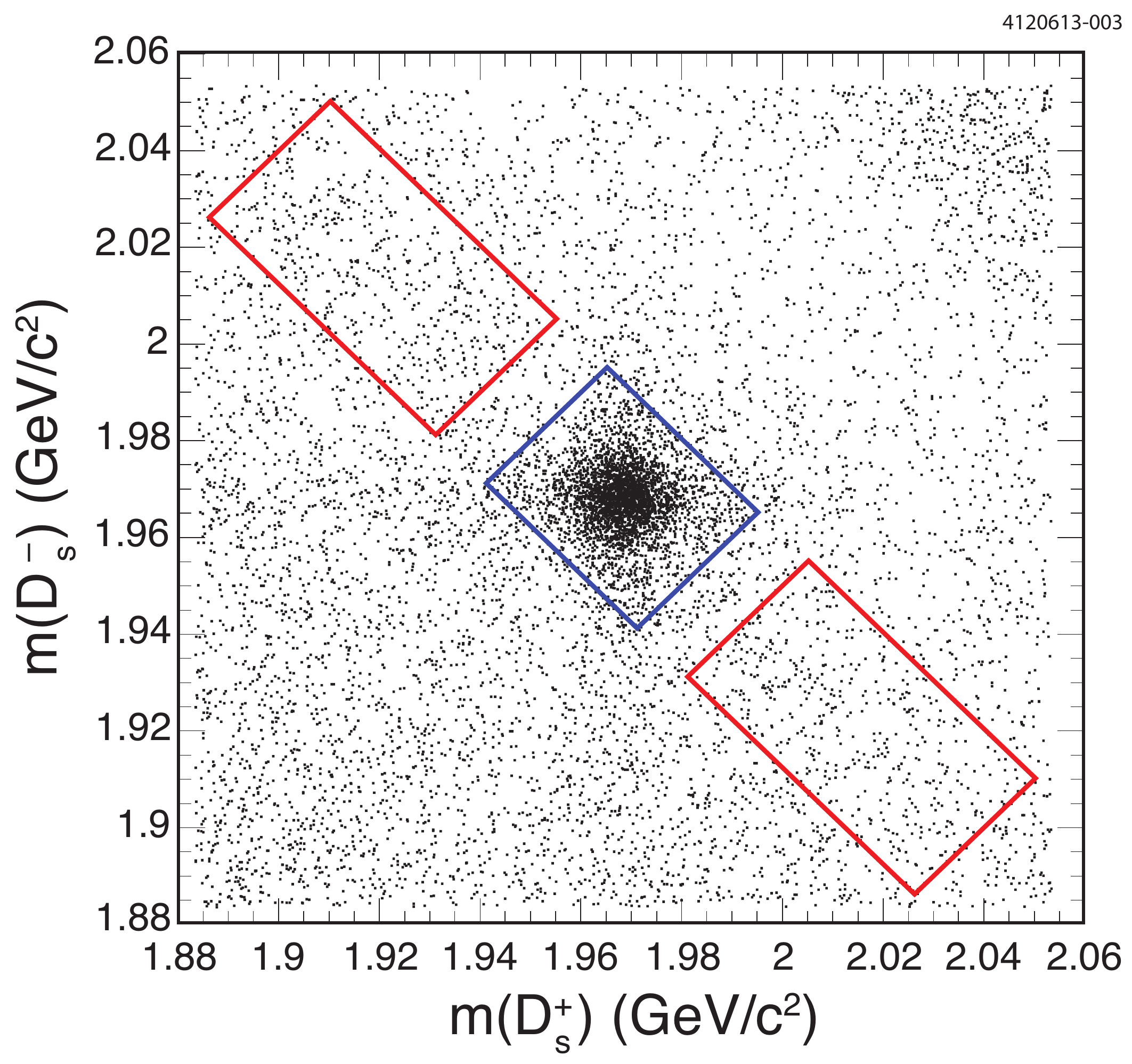}
 \caption{\label{fig:dt}(Color online) Invariant mass of \Dsm\ candidate versus invariant mass of \Dsp\ candidate for all double tag candidates (combining all 256 channels).  The central blue box indicates the signal region, while the two red boxes displaced along the diagonal indicate the sideband region.}
 \end{center}
\end{figure}

Double tag yields are determined by counting events in a signal region in the plane of $m(\Dsp)$ versus $m(\Dsm)$.  The signal region requires the mean invariant mass $\overline{m} \equiv (m(\Dsp) + m(\Dsm))/2$ to satisfy $|\overline{m} - 1.9682\textrm{ GeV}/c^2| < 12\textrm{ MeV}/c^2$ and the invariant mass difference $\Delta m \equiv m(\Dsp) - m(\Dsm)$ to satisfy $|\Delta m| < 30\textrm{ MeV}/c^2$.  Combinatoric backgrounds vary in $\overline{m}$ but are largely flat in $\Delta m$ for small values of that variable; hence we define a sideband region with the same $\overline{m}$ requirement but with $50 < |\Delta m| < 140\textrm{ MeV}/c^2$.  This sideband region has been found to model the signal region well for all but a few peaking backgrounds arising from open charm production.  The expected yields of these peaking backgrounds are again determined from generic MC, and form less than $1\%$ of the total double tag yield.  Fig.~\ref{fig:dt} shows the $m(\Dsp)$ vs.\ $m(\Dsm)$ distribution for all data double tag candidates, as well as the signal and sideband regions.

\if\usesections1\section{Branching Fraction Fit Procedure}\fi
We perform a maximum likelihood fit, with the branching fractions and $N_{\Dsstar\Ds}$ as parameters, to the observed single and double tag yields.  
The small expected crossfeeds and residual external peaking backgrounds to double tag modes are included in the fit.  Systematic uncertainties are propagated to the final results by altering the fit inputs accounting for appropriate correlations. 

We validate the self-consistency of the yield determinations, efficiencies, and branching fraction fit procedure on the generic MC sample, which corresponds to an integrated luminosity of 20 times the recorded dataset.  We reproduce the input parameters of the simulation with a $\chi^2/\mathrm{d.o.f.} = 19.1/17$ and conclude that the procedure has no significant inherent biases.  We also test the branching fraction fitter on pseudoexperiments with very small yields and find that it produces reasonable central values and pull distributions for the output parameters, even when many of the measured yields are in the low statistics regime.

\if\usesections1\section{Systematic Uncertainties and Cross-Checks}\fi

We check the time stability of the observed $\KS\Kp$ and $\Kp\Km\pip$ cross-sections, and the $\Dsp/\Dsm$ yield ratios in these modes.  No significant time dependence is seen.

We rely on signal lineshape parametrizations taken from signal MC when performing fits to obtain single tag yields in the data.  The actual lineshape might be different.  We perform a second set of fits allowing the widths of the signal peaks to vary by an overall mode-dependent scale factor.  The largest excursions are seen in modes with photons in the final state, where changes in yields up to 18.6\% are seen ($\KS\Kp\piz$).  We use the difference in the fixed-width and floating-width yields as uncertainties on the signal yields from the parametrizations.  Examining the eventual results of the branching fraction fit, we do not see evidence that these yield excursions are correlated, so these uncertainties are treated on a mode-by-mode basis, only correlated between \Dsp\ and \Dsm\ for a single mode.

For single tag yields, we subtract peaking backgrounds via the background shape in the yield fit.  For double tag yields, we explicitly subtract crossfeeds and external backgrounds.  The uncertainty due to the crossfeed estimates, and due to the single tag background parametrization, is obtained by refitting with the total open charm background estimates reduced by 20\%; the effect on the results is generally negligible, giving an uncertainty $<0.2\%$ on $\Br(\Ds\to\Km\Kp\pip)$.  The largest effect is 1.0\% on $\Br(\Ds\to\KS\KS\pip)$.

We independently compare the Monte Carlo and data rates for random pion pairs to be reconstructed as \KS\ candidates in modes with \KS\ daughters, by extracting the single tag yields in sidebands of \KS\ mass in data.  The generic MC overestimates the background, compared to data, and we correct for this effect.  The statistical uncertainty in the difference of the rates in data and generic MC is considered a systematic uncertainty.

We assign a number of systematic uncertainties to account for differences in predicted and actual efficiencies for reconstructing final state particles.  These are assumed to be fully correlated across all efficiencies.  We observe that for some neutral hadrons simulation overestimates the reconstruction efficiency, so we apply corrections of $-6.0$\% per \piz, $-6.5$\% per $\etagg$ candidate, and $-4.6$\% per $\etaprrg$ candidate. The uncertainties applied to the efficiencies are 0.3\% per charged pion, including \KS\ daughters; 0.4--2.5\%  for decays with charged kaons, where the exact value depends on the momentum spectra of the kaons; 0.9\% per \KS; 1.2--1.8\% per \piz, depending on the momentum spectra; 4.0\% per $\etagg$ candidate and $\etaprrg$ candidate; and an additional 4.0\% for $\etaprrg$ candidates, added in quadrature with the previous uncertainty.  Similarly we correct the simulation for observed small momentum-dependent differences between simulation and data for particle ID efficiency, and we assign an 
uncertainty of 0.2\% per pion and 0.3\% per kaon, correlated for all efficiencies.

We use certain intermediate particle decays ($\KS \to \pip\pim$, $\eta \to \gamma\gamma$, $\eta \to \pip\pim\piz$, $\eta' \to \pip\pim \eta$, and $\eta' \to \pip\pim\gamma$), which themselves have uncertainties in their branching fractions.  We correct our simulations to the PDG 2012 \cite{Beringer:1900zz} world averages for these branching fractions and include systematic uncertainties of 0.07\%, 0.7\%, 1.2\%, 1.6\%, and 2.0\%, respectively.

The predicted reconstruction efficiencies for various \Ds\ decays depend on the resonant substructure of the decays, as these determine the momentum spectra of observed final state particles.  Differences between data and simulation in decays to three or more final state particles may therefore bias the efficiency.  We obtain a measurement of the efficiency from data and compare to the simulation to determine the potential size of such biases.  First, we parametrize the efficiency as a function of ``Dalitz variables,'' the invariant mass squareds $m_{ab}^2$ of pairs of final state particles.  This parametrization is done with a multilayer perceptron neural network from the TMVA package \cite{Hocker:2007ht}, used as a regression tool.  It is trained on simulation samples where generated events which are reconstructed in simulation are assigned the value 1, and those which are not reconstructed are assigned the value 0; with the appropriate choice of estimator the neural network output value converges to the local value of the 
efficiency at 
each point in phase space.  This procedure can be applied to simulated samples where the generated events are not uniformly distributed in phase space.  We then use our likelihood fits to data to obtain per-event signal weights using the sPlot procedure \cite{Pivk:2004ty}, which essentially functions like a sideband subtraction technique using all available events.  Having obtained the efficiency as a function of position in phase space and a background-subtracted model of the distribution of data events in the phase space, the weighted harmonic mean of the expected efficiencies $\sum w_i/\sum (w_i/\epsilon_i)$ provides the overall efficiency, as estimated from data.  We can perform this procedure in generic MC as well, which gives an estimate of the bias in the determined efficiency caused by the presence of background.  We generally assign the quadrature sum of the departures from unity of the data/signal MC and generic MC/signal MC ratios as the uncertainty due to the resonant substructure for a mode.  
This ranges from 0.6\% for \Km\Kp\pip\ to 9.0\% for \Km\Kp\pip\piz.  In the \pip\pip\pim\ mode, there is evidence that a contribution not modeled in the simulation is present in data.  In this case we correct the efficiency determined from simulation by 5\%, to match the value estimated from data, and apply an uncertainty of 2\%.  Additional uncertainties of (0.6--0.7)\% are applied to account for the fraction of events rejected by \KS\ vetoes.  Finally, we see an excess of events in data for $\pip\piz\eta$ that have $m(\pip\piz)$ above the upper bound for our $\rho^+$ selection, compared to MC simulation.  We find that the MC-determined ratio of yields after our $m(\pip\piz)$ selection to the full phase space is low by $(-13 \pm 2)\%$ and correct our efficiency to reflect this, as our final result is the branching fraction for the full $m(\pip\piz)$ phase space.
\begin{table*}[!t]
\caption{\label{tbl:bfresults}Results of the fit for \Ds decay branching fractions; comparison to the PDG 2012 fit result; ratio to $\Br(\Ds\to\Km\Kp\pip)$ for this result; and $CP$ asymmetries $\mathcal{A}_{CP}$ for this result. For \cleoc results uncertainties are statistical and systematic, respectively; for the PDG fit total uncertainties are shown.  For PDG results with a \dag\ indication, we show the result for $\rho^+ X$ rather than $\pip\piz X$ as the latter is unavailable. }
\begin{center}
\begin{tabular}{lcccc}
\hline\hline
Mode & This result $\Br$ (\%) & PDG 2012 fit $\Br$ (\%) & $\Br/\Br(\Ds\to\Km\Kp\pip)$ & $\mathcal{A}_{CP}$ \\
\hline
$\KS\Kp$ & $1.52 \pm 0.05 \pm 0.03$ &  $1.48 \pm 0.08$ & $0.274 \pm 0.006 \pm 0.005$ & $+0.026 \pm 0.015 \pm 0.006$\\
$\Km\Kp\pip$ & $5.55 \pm 0.14 \pm 0.13$ &  $5.49 \pm 0.27$ & 1 & $-0.005 \pm 0.008 \pm 0.004$ \\
$\KS\Kp\piz$ & $1.52 \pm 0.09 \pm 0.20$  & --- & $0.274 \pm 0.016 \pm 0.037$ & $-0.016 \pm 0.060 \pm 0.011$\\
$\KS\KS\pip$ & $0.77 \pm 0.05 \pm 0.03$  & --- & $0.138 \pm 0.009 \pm 0.006$ & $+0.031 \pm 0.052 \pm 0.006$\\
$\Km\Kp\pip\piz$ & $6.37 \pm 0.21 \pm 0.56$  & $5.6\pm 0.5$ & $1.147 \pm 0.034 \pm 0.099$ & $+0.000 \pm 0.027 \pm 0.012$\\
$\KS\Kp\pip\pim$ & $1.03 \pm 0.06 \pm 0.08$ & $0.96 \pm 0.13$ & $0.185 \pm 0.011 \pm 0.015$ & $-0.057 \pm 0.053 \pm 0.009$\\
$\KS\Km\pip\pip$ & $1.69 \pm 0.07 \pm 0.08$ & $1.64 \pm 0.12$ & $0.304 \pm 0.010 \pm 0.014$ & $+0.041 \pm 0.027 \pm 0.009$\\
$\pip\pip\pim$ & $1.11 \pm 0.04 \pm 0.04$  & $1.10 \pm 0.06$ & $0.200 \pm 0.007 \pm 0.008$  & $-0.007 \pm 0.030 \pm 0.006$\\
$\pip\eta$ combined & $1.67 \pm 0.08 \pm 0.06$ & $1.83 \pm 0.15$ & $0.301 \pm 0.014 \pm 0.013$ & $+0.011 \pm 0.030 \pm 0.008$ \\
\hspace{1.5em}$\pip\etagg$ & $1.75 \pm 0.08 \pm 0.16$ & --- & $0.315 \pm 0.013 \pm 0.031$ & $+0.006 \pm 0.036 \pm 0.009$\\
\hspace{1.5em}$\pip\etatpi$ & $1.63 \pm 0.12 \pm 0.06$ & ---  & $0.294 \pm 0.020 \pm 0.011$ & $+0.024 \pm 0.054 \pm 0.016$\\
$\pip\piz\eta$ & $9.2 \pm 0.4 \pm 1.1$ & $8.9 \pm 0.8$ \dag & $1.66 \pm 0.07 \pm 0.21$  & $-0.005 \pm 0.039 \pm 0.020$\\
$\pip\eta'$ combined & $3.94 \pm 0.15 \pm 0.20$ & $3.94 \pm 0.33$ & $0.709 \pm 0.025 \pm 0.039$  & $-0.022 \pm 0.022 \pm 0.006$ \\
\hspace{1.5em}$\pip\etaprgg$ & $4.07 \pm 0.17 \pm 0.30$ & ---  & $0.73 \pm 0.03 \pm 0.06$ & $-0.052 \pm 0.027 \pm 0.008$\\
\hspace{1.5em}$\pip\etaprtpi$ & $3.7 \pm 0.5 \pm 0.2$  & --- & $0.68 \pm 0.08 \pm 0.04$  & $+0.011 \pm 0.097 \pm 0.032$\\
\hspace{1.5em}$\pip\etaprrg$ & $3.91 \pm 0.17 \pm 0.33$ & --- & $0.70 \pm 0.03 \pm 0.06$  & $+0.031 \pm 0.039 \pm 0.007$ \\
$\pip\piz\eta'$ & $5.6 \pm 0.5 \pm 0.6$  & $12.5 \pm 2.2$ \dag & $1.01 \pm 0.08 \pm 0.12$  & $-0.004 \pm 0.074 \pm 0.019$ \\
$\Kp\pip\pim$ & $0.654 \pm 0.033 \pm 0.025$ & $0.69 \pm 0.05$ & $0.118 \pm 0.006 \pm 0.005$  & $+0.045 \pm 0.048 \pm 0.006$ \\ 
\hline\hline
\end{tabular}
\end{center}
\end{table*}

We use tight cuts on \mrec\ in order to reduce backgrounds for most single tag yields.  The efficiency of this cut depends on the momentum spectrum of the \Ds\ mesons produced via $e^+e^- \to \Ds^*\Ds$.  We consider two effects that might alter this.  First, initial state photon radiation can result in a true center of mass energy lower than the nominal one; this shifts \mrec\ towards higher values and can cause events to fail the tight cut.  Secondly, indirect \Ds\ mesons produced in $\Ds^* \to \piz\Ds$ are much more likely to be accepted by the tight \mrec\ cut than those from $\Ds^* \to \gamma\Ds$, so uncertainties in $\Br(\Ds^* \to \piz\Ds)$ translate to uncertainties in the tight \mrec\ cut efficiency.  We measure the ratio of observed single tag \Km\Kp\pip\ yields for the tight and loose \mrec\ cuts in data and signal MC, which differ by $(0.4 \pm 0.6)\%$.  In addition we find that changing $\Br(\Ds^* \to \piz\Ds)$ by the uncertainty on the PDG average\ and allowing a 0.7\% contribution from $\Br(\Ds^* 
\to e^+e^-\Ds)$ \cite{CroninHennessy:2011xp} causes excursions of 0.3\% on the efficiency.  Combining these effects we add systematic uncertainties of 0.6\% and 0.3\% in quadrature, correlated for single tag efficiencies in modes with tight \mrec\ cuts.

We allow only one candidate per reconstructed final state per event, and there is some inefficiency associated with this choice; if the rate of events with multiple candidates differs in data and in MC, our nominal efficiencies will be in error.  We estimate the size of such an effect by computing the ratio of \Ds\ yield in the {\em rejected} candidates to the yield in the {\em chosen} candidates.  We see agreement in the rates between data and MC within the statistical uncertainty.  The difference in the central values of the ratios is taken as a systematic uncertainty.

Final state photon radiation (FSR) from charged \Ds\ daughters is modeled in simulation using the \textsc{Photos} package, which allows interference between the radition from different daughters.  Events with significant FSR will have low reconstructed \Ds\ candidate mass and will have lower efficiency.  We determine the difference in efficiency between simulated events where \textsc{Photos} does not generate a FSR photon and the inclusive sample, and assign 30\% of this difference as a systematic uncertainty.  The largest value is 1.4\% for $\pip\pip\pim$.

\if\usesections1\section{Results and Discussion}\fi
The results of the branching fraction fit and $CP$ asymmetry analysis are shown in Table~\ref{tbl:bfresults}; the correlation matrix is available in the Supplemental Material \cite{supplemental}.  
The statistical uncertainty on $\Br(\Ds \to \Km\Kp\pip)$ is 2.5\%, and the quadrature sum of the statistical and systematic uncertainty is 3.4\%.  This compares to the PDG 2012 fit uncertainty of 4.9\% \cite{Beringer:1900zz}.  The PDG 2012 fit includes previous CLEO-c results and is therefore correlated with this measurement.  The largest single contribution to the \Km\Kp\pip\ systematic uncertainty is the kaon tracking efficiency, which is 1.5\%; the subleading contributors are the single tag lineshape uncertainties and the particle ID uncertainty.  In addition we obtain $N_{\Ds^*\Ds} = (5.67 \pm
0.15 \textnormal{(stat)} \pm 0.10 \textnormal{(syst)}) \times 10^5$; this gives $\sigma_{\Ds^*\Ds}(4.170\textnormal{ GeV}) = 0.967 \pm 0.026 \textnormal{(stat)} \pm 0.017 \textnormal{(syst)} \pm 0.010 \textnormal{(lum)}\textnormal{ nb}$.  The luminosity normalization is derived using the procedure discussed in Ref.~\cite{Dobbs:2007ab}.  No notable $CP$ asymmetries are found, the most significant being 1.6$\sigma$ in $\KS\Kp$.

Compared to our previous result based on 298 \invpb\ of data, all values are consistent with the exception of $\Km\Kp\pip\piz$, which has increased 13\%.  The change is due to improvements in our understanding of the resonant substructure and our \piz\ reconstruction efficiency.

We find $\Br(\Ds \to \pip\piz\eta')$ to be less than half the PDG 2012 value of $\Br(\Ds \to \rho^+\eta')$, which is set by a branching ratio measured in Ref.~\cite{Jessop:1998bc} ($\Br(\Ds \to \rho^+\eta')/\Br(\Ds \to \phi\pip) = 2.78 \pm 0.28 \pm 0.30$).  The PDG value causes a large tension between the inclusive measurement of $\Br(\Ds \to \eta' X)$ = $(11.7 \pm 1.8)$\% \cite{Dobbs:2009aa} and the sum of known exclusive branching fractions $(18.6 \pm 2.3)$\%. With our new $\Br(\Ds \to \pip\eta')$ and $\Br(\Ds \to \pip\piz\eta')$, and using the PDG fits for $\Br(\Ds \to \Kp\eta')$ and $\Br(\Ds \to \eta'e^+\nu)$, we find the sum of exclusive decays involving $\eta'$ to be $(11.7 \pm 0.9)$\%, in very good agreement with the inclusive determination.

These results supersede previous \cleoc determinations of $\Ds$ branching fractions and $CP$ asymmetries.  However they do not supersede the measurement of $\Br(\Ds\to\rho^+\eta) = (8.9 \pm 0.6 \pm 0.5)\%$ from Ref.~\cite{Naik:2009tk}, as that measurement explicitly looks only at the $\rho^+$ contribution instead of the full $\pip\piz$ phase space.

\if\usesections1\section{Conclusion}\fi
We have measured the absolute branching fractions for thirteen \Ds\ decays, reconstructed in sixteen final states, using a double tag technique.  This provides the most precise available values of the reference branching fractions $\Br(\Ds \to \Km\Kp\pip)$ and $\Br(\Ds \to \KS\Kp)$.  The thirteen decays together form ($40.7 \pm 1.8)$\% of \Ds\ decays.  No evidence of direct $CP$ violation was found.  We find that $\Br(\Ds \to \pip\piz\eta')$ is significantly smaller than the current world average, and our measured value resolves the tension between the inclusive and exclusive determinations of $\Br(\Ds \to \eta' X)$.  Finally we have also determined the cross-section for $e^+ e^- \to \Ds^* \Ds$ at $E_\com = 4.17$ GeV to be $\sigma_{\Ds^*\Ds}(4.170\textnormal{ GeV}) = 0.967 \pm 0.026 \textnormal{(stat)} \pm 0.017 \textnormal{(syst)} \pm 0.010 \textnormal{(lum)}\textnormal{ nb}$.

\if\usesections1\acknowledgments\fi

We gratefully acknowledge the effort of the CESR staff
in providing us with excellent luminosity and running conditions.
This work was supported by
the A.P.~Sloan Foundation,
the National Science Foundation,
the U.S. Department of Energy,
the Natural Sciences and Engineering Research Council of Canada, and
the U.K. Science and Technology Facilities Council.
\bibliographystyle{apsrev4-1}
\
\bibliography{dspaper}

\end{document}